\documentclass[superscriptaddress,preprintnumbers,twocolumn,amsmath,amssymb]{revtex4-1}

\usepackage{graphicx}
\usepackage{dcolumn}
\usepackage{bm}
\usepackage{here}

\begin{document}

\preprint{Journal-ref: Scripta Materialia {\bf 223}, 115081 (2023)}

\title{Direct observation of multiple conduction-band minima in high-performance thermoelectric SnSe}

\author{Mario~Okawa}
\affiliation{Department of Applied Physics, Waseda University, Shinjuku, Tokyo 169-8555, Japan}
\author{Yuka~Akabane}
\affiliation{Department of Applied Physics, Waseda University, Shinjuku, Tokyo 169-8555, Japan}
\author{Mizuki~Maeda}
\affiliation{Department of Applied Physics, Waseda University, Shinjuku, Tokyo 169-8555, Japan}
\author{Gangjian~Tan}
\altaffiliation[Present address: ]{State Key Laboratory of  Advanced Technology for Materials Synthesis and Processing,
	Wuhan University of Technology, Wuhan 430070, China}
\affiliation{Department of Chemistry, Northwestern University, Evanston, IL 60208, USA}	
\author{Li-Dong~Zhao}
\altaffiliation[Present address: ]{School of Materials Science and Engineering, Beihang University, Beijing 100191, China}
\affiliation{Department of Chemistry, Northwestern University, Evanston, IL 60208, USA}
\author{Mercouri~G.~Kanatzidis}
\affiliation{Department of Chemistry, Northwestern University, Evanston, IL 60208, USA}
\affiliation{Materials Science Division, Argonne National Laboratory, Lemont, IL 60439, USA}
\author{Takeshi~Suzuki}
\affiliation{Institute for Solid State Physics, University of Tokyo, Kashiwa, Chiba 277-8581, Japan}
\author{Mari~Watanabe}
\altaffiliation[Present address: ]{National Metrology Institute of Japan, AIST, Tsukuba, Ibaraki 305-8569, Japan}
\affiliation{Institute for Solid State Physics, University of Tokyo, Kashiwa, Chiba 277-8581, Japan}
\author{Jiadi~Xu}
\affiliation{Institute for Solid State Physics, University of Tokyo, Kashiwa, Chiba 277-8581, Japan}
\author{Qianhui~Ren}
\affiliation{Institute for Solid State Physics, University of Tokyo, Kashiwa, Chiba 277-8581, Japan}
\author{Masami~Fujisawa}
\affiliation{Institute for Solid State Physics, University of Tokyo, Kashiwa, Chiba 277-8581, Japan}
\author{Teruto~Kanai}
\affiliation{Institute for Solid State Physics, University of Tokyo, Kashiwa, Chiba 277-8581, Japan}
\author{Jiro~Itatani}
\affiliation{Institute for Solid State Physics, University of Tokyo, Kashiwa, Chiba 277-8581, Japan}
\author{Shik~Shin}
\affiliation{Institute for Solid State Physics, University of Tokyo, Kashiwa, Chiba 277-8581, Japan}
\affiliation{Materials Innovation Research Center, University of Tokyo, Kashiwa, Chiba 277-8561, Japan}
\affiliation{Office of University Professor, University of Tokyo, Kashiwa, Chiba 277-8581, Japan}
\author{Kozo~Okazaki}
\affiliation{Institute for Solid State Physics, University of Tokyo, Kashiwa, Chiba 277-8581, Japan}
\affiliation{Materials Innovation Research Center, University of Tokyo, Kashiwa, Chiba 277-8561, Japan}
\affiliation{Trans-scale Quantum Science Institute, University of Tokyo, Bunkyo, Tokyo 113-0033, Japan}
\author{Naurang~L.~Saini}
\affiliation{Dipartimento di Fisica, Universit\`{a} degli Studi di Roma ``La Sapienza,'' 00185 Rome, Italy}
\author{Takashi~Mizokawa}
\affiliation{Department of Applied Physics, Waseda University, Shinjuku, Tokyo 169-8555, Japan}

\date{\today}

\begin{abstract}
We report time- and angle-resolved photoemission spectroscopy on SnSe which currently attracts great interest
due to its extremely high thermoelectric performance.
Laser-assisted photoemission signals are observed within $\pm$20 fs of the pump pulse arrival.
Around 30--50 fs after the photoexcitation, the conduction band minima are not populated by the photoexcited electrons
while the valence bands are considerably broadened.
In going from 90 fs to 550 fs after the photoexcitation, the photoexcited carriers are decayed into the multiple conduction
band minima.
The observed conduction bands are consistent with the band structure calculations.
The multiple conduction minima suggest possibility of high and anisotropic thermoelectric performance of $n$-type SnSe single crystal if it is realized.
\end{abstract}

\maketitle

Thermoelectric power generation is one of the key technologies for sustainable and secure energy supply.
Among the various candidate materials for thermoelectric power generation devices, the multiband IV-VI semiconductors such as Pb$Q$ ($Q$=Se, Te) \cite{Heremans2008,Pei2011,Pei2012,Biswas2012,Wu2014} and Sn$Q$ ($Q$=Se, Te) \cite{Zhao2014,Tan2014} have been attracting great interest due to their high thermoelectric performance with relatively good electrical conductivity, enhanced Seebeck coefficient, and strongly suppressed lattice thermal conductivity.
The large Seebeck effect is derived from the multiple valence band maxima and the conduction band minima in the multiband band structure.
The suppression of lattice thermal conductivity is associated with the lattice disorder and anharmonic dynamics which are derived from the lone pairs due to the Pb(Sn) $s$ and $p$ orbitals \cite{Li2015}.
In SnSe, it has been proposed that the small energy separation between the multiple valence band maxima plays essential roles for the enhancement of Seebeck coefficient \cite{Zhao2014}.
Interplay between the lone pair effect and the multiband electronic structure provides the unique electron-lattice coupling to the Pb$Q$ and Sn$Q$ systems.
It is expected that the specific electron lattice coupling manifests itself in their ultrafast optical response. In this context, it would be interesting to investigate optically induced electronic and lattice dynamics of the thermoelectric Pb$Q$ and Sn$Q$ systems by means of time-resolved spectroscopy.
In particular, time- and angle-resolved photoemission spectroscopy (TARPES) enables us to probe dynamics of photoexcited carriers and temporal evolution of the band structure.

\begin{figure}[!b]
\centering
\includegraphics[width=7.5cm]{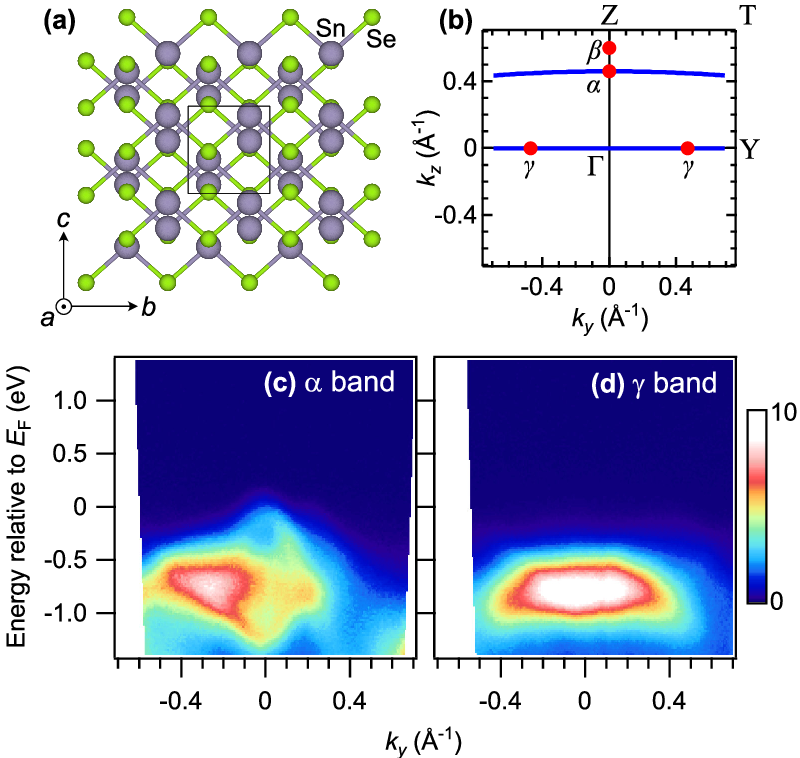}
\caption{
(a) The \textit{Pnma} crystal structure of SnSe created by VESTA \cite{VESTA}. (b) The Brillouin zone of SnSe. $k_y$ ($k_z$) represents the wave number along the $\Gamma$Y ($\Gamma$Z) direction. The circles correspond to the valence band maximum points of the $\alpha$, $\beta$, and $\gamma$ bands.
Photoemission intensity distributions as functions of energy and $k_y$ before photoexcitation for (c) $\alpha$ band cut of SnSe, (d) $\gamma$ band cut of SnSe, whose momentum locations are shown by the solid lines in (b).
}
\end{figure}

\begin{figure*}[!t]
\centering
\includegraphics[width=17.8cm]{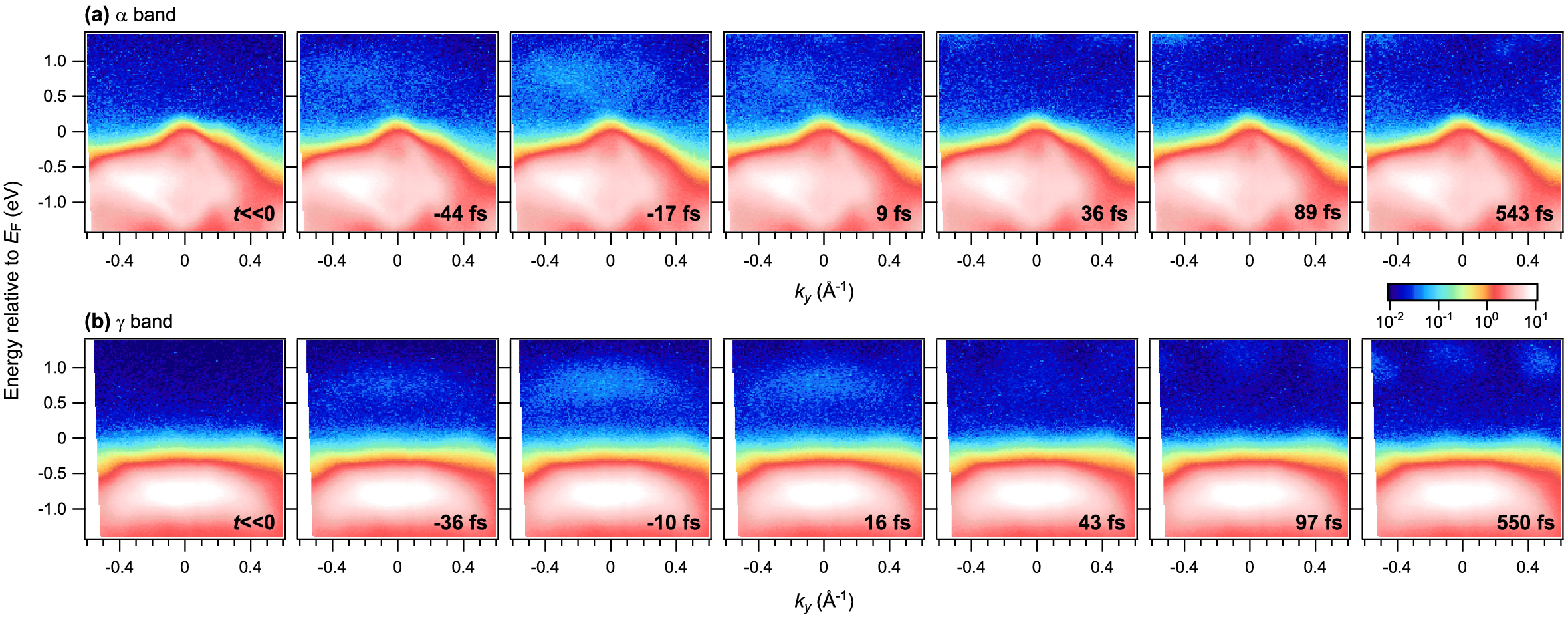}
\caption{
Time evolution after photoexcitation of the photoemission intensity distributions as functions of energy and $k_y$ for (a) $\alpha$ band cut and (b) $\gamma$ band cut in SnSe. The photoemission intensities are plotted in the logarithmic scale in order to emphasize the photo-induced change.
}
\end{figure*}

SnSe has a layered structure constructed with strong bonds along the $bc$ plane and weak bonds along the $a$-axis as shown in Fig.\ 1(a) \cite{Chattopadhyay1986}.
Since the discovery of ultrahigh thermoelectric performance in Na-doped SnSe \cite{Zhao2016}, the electronic structure of SnSe and Na-doped SnSe with the multiple valence band maxima has been extensively studied by means of band structure calculations and angle-resoled photoemission spectroscopy \cite{Lu2017,Wang2017,Nagayama2017,Tayari2018,Maeda2018,Pletikosic2018,Wang2018,Wu2018}.
The observed valence band structure is basically consistent with the theoretical predictions as well as the transport properties \cite{Zhao2016,Car1978,Makinistian2009,Kutorasinski2015,Shi2015}.
Compared to the valence band, the nature of the conduction band of SnSe has not been studied so far.
In addition, dynamics of photoexcited carriers may provide useful information for understanding of the unusual electronic and lattice properties of SnSe. In the present work, we report a TARPES study on SnSe. The present results provide interesting insights on the ultrafast dynamics of the photoexcited electrons of SnSe in addition to information on the conduction bands. The observed conduction band minima are consistent with the prediction of density functional calculations, suggesting possibility of high thermoelectric performance due to the multiple conduction band minima for $n$-type SnSe if it is realized.


Single crystals of SnSe were grown as reported in the literature \cite{Zhao2014,Zhao2016}.
The crystals were well characterized using x-ray diffraction, electron microscopy, and transport measurements as described in Ref.\ \cite{Zhao2014}.
TARPES experiments have been performed at LASOR, Institute for Solid State Physics, University of Tokyo \cite{Suzuki2021}.
The base pressure of the spectrometer was in the $9 \times 10^{-11}$ Pa.
The samples were cleaved and measured at 80 K.
The energy of the Ti:sapphire laser was 1.55 eV and the energy of the probe light was 21.7 eV. The repetition rate was set to 10 kHz, and the fluence of the incident pump was set to $\sim$0.95 mJ/cm$^2$. The pump and probe photons were $p$-polarized relative to the sample surface. The photoelectrons were detected by using a hemispherical electron analyzer (Omicron-Scienta R4000). The total energy resolution was 227 meV. The pulse duration of the fundamental laser was 35 fs. Local density approximation (LDA) calculations were performed by {\sc Quantum} ESPRESSO 5.30 \cite{QE1,QE2}.
Pseudopotentials of {\tt Sn.pz-dn-rrkjus\_psl.0.2.UPF} and {\tt Se.pz-n-rrkjus\_psl.0.2.UPF} were used for the calculations.


Figures 1(c) and (d) show the photoemission intensity distributions as functions of energy and $k_y$ for SnSe before the pump pulse arrives. Here, $k_y$ represents the wave number along the $\Gamma$Y direction. The valence band maxima near the Z point are included in the cut of Fig.\ 1(c) ($\alpha$ band cut). The $\gamma$ band cut shown in Fig.\ 1(d) includes the zone center and the valence band maxima near the Y point. These results are consistent with the previous ARPES results \cite{Lu2017,Wang2017,Nagayama2017,Tayari2018,Maeda2018,Pletikosic2018,Wang2018,Wu2018}.

Figure 2(a) shows the relaxation process after the pump pulse arrival for the photoemission intensity distribution along the $\alpha$ band cut of SnSe. In order to emphasize the photoinduced change, the logarithm of the photoemission intensity is plotted.  Within $\pm$20 fs of the pump pulse arrival, a downwardly convex parabolic signal is observed just above the valence band maxima. The intensity distribution of the photoinduced signal resembles that of the valence band. The energy difference between them is $\sim$1.55 eV which corresponds to the photon energy of pump pulse. This indicates that the photoinduced signal is due to laser-assisted photoemission (LAPE) process \cite{LAPE1,LAPE2,LAPE3,LAPE4,Rohwer2011}.
This assignment is consistent with the $p$-polarized pump pulse which can enhance the LAPE effect \cite{LAPE4}.
The reference (0 fs) of the delay time is determined by the peak position in the temporal evolution of the LAPE signal.
At 36 fs after the photoexcitation, the LAPE signal is almost depleted.

As for the $\gamma$ band cut including the $\Gamma$ point [Fig.\ 2(b)], a flat signal is observed around 0.5 eV above the Fermi level within $\pm$20 fs of the pump pulse arrival. The intensity distribution of the photoinduced signal also resembles that of the valence band, and the energy difference between them is $\sim$1.55 eV. Therefore, the photoinduced signal can be assigned to LAPE. At 43 fs after the photoexcitation, the LAPE signal is almost depleted. Interestingly, at 97 fs and 550 fs, residual populations of the conduction bands are observed at three points along the cut. The three points may correspond to the three local minima of the conduction bands predicted for the $Pnma$ phase of SnSe.

\begin{figure}[!t]
\centering
\includegraphics[width=8cm]{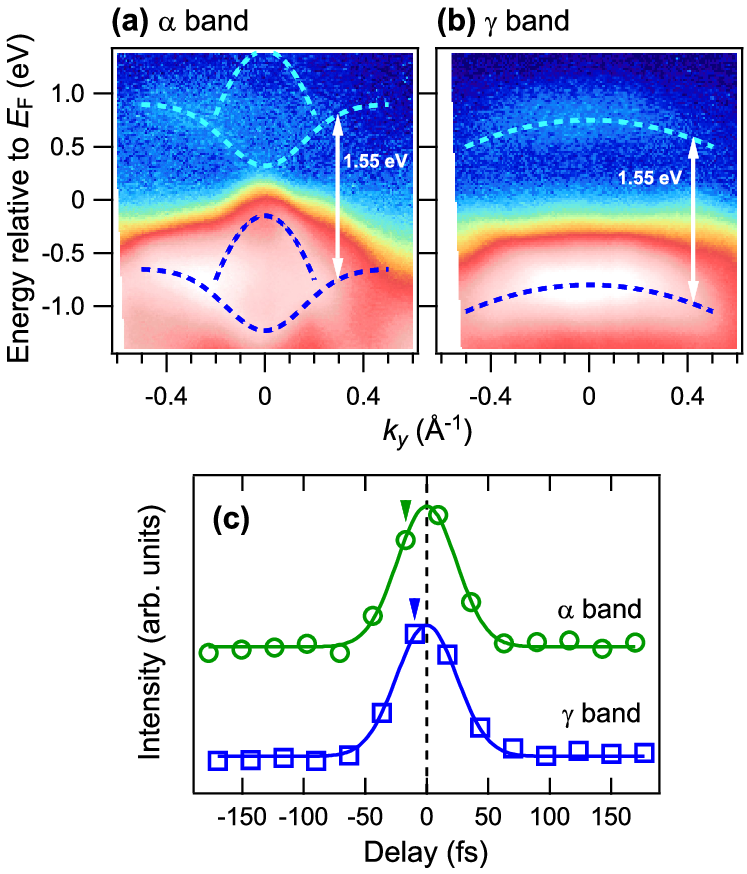}
\caption{
LAPE signals at (a) $-17$ fs for $\alpha$ band cut and (b) $-9$ fs for $\gamma$ band cut, which correspond to triangles in the panel (c). The dashed lines are guides to eyes for the valence band structure and these LAPE signals.
(c) Temporal evolution of the integrated photoemission intensity (0.5--1 eV) for the $\alpha$ band cut and $\gamma$ band cut in SnSe. The solid lines are fitting results using the Gaussian functions.}
\end{figure}

Figures 3(a) and (b) show the LAPE signals for $\alpha$ band cut and $\gamma$ band cut respectively. As discussed in the previous paragraphs, the energy difference between the LAPE signals and the valence band signals is 1.55 eV. Figure 3(c) shows temporal evolution of the integrated intensity between the energy of 0.5 eV and 1 eV for $\alpha$ band cut and $\gamma$ band cut. The LAPE peaks in the temporal evolution are fitted to Gaussian functions. The peak positions of the fitted Gaussians are assigned to the reference (0 fs) of the delay time.

\begin{figure}[!t]
\centering
\includegraphics[width=8.6cm]{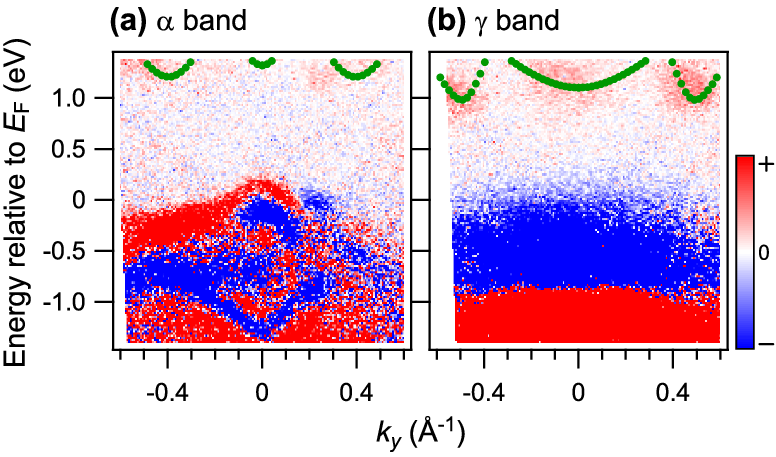}
\caption{
Difference of photoemission intensity distributions between before and after ($\sim$550 fs) the photoexcitation for (a) $\alpha$ band cut and (b) $\gamma$ band cut in SnSe. The green dots indicate calculated band dispersions.}
\end{figure}

Figures 4(a) and (b) show the difference spectra obtained by subtracting the intensity distribution before the pump pulse arrival from those after the arrival (543 fs for $\alpha$ band cut and 550 fs for $\gamma$ band cut). For the $\alpha$ band cut in Fig.\ 4(a), photoinduced electrons are observed around 1.3 eV above the Fermi level. In the $\gamma$ band cut in Fig.\ 4(b), they are more clearly observed around 1.0 eV above the Fermi level. Interestingly, there are three segments of the photoinduced spectral weight in the $\gamma$ band cut.
In addition to the photoinduced electrons in the conduction band, the valence band is considerably broadened involving the conduction band due to the photoexcitation.
The intensity distributions above the Fermi level are compared with the LDA band structure calculations which are shifted upwards in order to adjust the conduction band bottom. In the $\gamma$ band cut, the three segments correspond to the three local minima of the conduction band.

Around 40 fs after the pump pulse arrival in the $\gamma$ band cut, there are residual signals within the band gap ranging from the Fermi level to 1.0 eV (see Fig.\ 2). On the other hand, around 500 fs after the pump pulse arrival, the in-gap spectral weight is completely suppressed. The photoinduced spectral weight below the conduction band bottom resembles that observed in Ta$_2$NiSe$_5$ \cite{Okazaki2018}, which is considered to be a candidate of excitonic insulator \cite{Wakisaka2009,Wakisaka2012, Seki2014,Lu2017_TNS,Mor2017}.
In the photoinduced gap collapse of Ta$_2$NiSe$_5$, the photoexcited carriers screen the Coulomb interaction between the valence band hole and the conduction band electron which provides the insulating ground state. A similar mechanism is proposed for another excitonic insulator candidate TiSe$_2$.\cite{Rohwer2011}
As for SnSe, the photoinduced in-gap state may indicate electron-hole correlation effect on the formation of its band gap. However, there is no indication of excitonic instability in SnSe. Here, we speculate that the strong Sn 5$s,p$-Se 4$p$ hybridization with the lone pairs formation \cite{Li2015} is destroyed by the photoexcited carriers. If the Sn 5$s,p$-Se 4$p$ hybridization is weakened and the chemical bonds in the Sn-Se polyhedral are suppressed, then the band gap is expected to be reduced or collapsed. For example, in the GW calculation for the $Cmcm$ phase with stacked rocksalt layers (without the distortion by the lone pairs), the magnitude of the band gap is predicted to be 0.464 eV \cite{Shi2015}.
On the other hand, it is calculated to be 0.829 eV for the $Pnma$ phase with the distortion due to the lone pairs \cite{Shi2015}.
Therefore, it can be deduced that the residual spectral weight in the band gap is associated with the photoinduced suppression of the lone pairs.

The $\alpha$ band and $\gamma$ band cuts in Fig.\ 4 are consistent with the calculations, confirming the anisotropy between the $b$ and $c$ axes.
The experimental confirmation of the conduction-band anisotropy between the $b$ and $c$ axes using TARPES provides valuable information for understanding electrical properties in the $n$-type single crystals.
However, the TARPES experiment with the fixed photon energy cannot observe the band dispersion along $a$ axis that is perpendicular to the cleaved surface.
The anisotropy between the $b$ and $c$ axes indicates that the $n$-type SnSe single crystal would show anisotropic thermoelectric performance if it is realized.
The present observation of the multiple conduction band minima suggests the high thermoelectric performance of $n$-type SnSe due to the multiple conduction band minima.
Actually, high-performance thermoelectricity with $ZT \gtrsim 1$ at $\sim$800 K was reported in recent studies on $n$-type electron-doped SnSe polycrystals and nanocrystals \cite{Li2018,Shang2019,Gainza2020,Cai2020,Su2021}.
Besides, the electronic structure including the band-gap and conduction-band anisotropy in $n$-type semiconductors (e.g., ZrNiSn \cite{Fu2020}) can be uncovered using ARPES.
In the future, ARPES of single-crystalline electron-doped SnSe is desired to design high-performance thermoelectric $n$-type SnSe comparable to $p$-type one.

In conclusion, we have performed TARPES on SnSe with extremely high thermoelectric performance in order to investigate their conduction bands as well as ultrafast dynamics of photoexcited electrons. The valence bands of SnSe are considerably broadened due to the photoexcitation. Around 40 fs after the pump pulse arrival, the conduction bands are not populated by the photoexcited electrons and there are residual signals within the band gap. In going from $\sim$90 fs to $\sim$550 fs, the photoexcited carriers are decayed into the multiple conduction band minima. The observed conduction bands are consistent with the band structure calculations. This observation suggests high thermoelectric performance of $n$-type SnSe due to the multiple conduction band minima.

\begin{acknowledgments}
At Tokyo, this work was supported by the Grants-in-Aid from the Japan Society of the Promotion of Science (JSPS) (No.\ 19H00659). At Argonne, this work was supported by the U.S.\ Department of Energy, Office of Science, Basic Energy Sciences, Materials Sciences and Engineering Division (materials synthesis).
This work was carried out under the Visiting Researcher's Program of the Institute for Solid State Physics,  University of Tokyo (Proposal No.\ B219).
\end{acknowledgments}

\end{document}